# Synchrotron powder X-ray diffraction and structural analysis of $Eu_{0.5}La_{0.5}FBiS_{2-x}Se_x$

K. Nagasaka[1], G. Jinno[1], O. Miura[1], A. Miura[2], C. Moriyoshi[3], Y. Kuroiwa[3], Y. Mizuguchi[1]*

1. Department of Electrical and Electronic Engineering, Tokyo Metropolitan University, 1-1, Minami-osawa, Hachioji 192-0397, Japan.

2. Faculty of Engineering, Hokkaido University, Kita-13, Nishi-8, Kita-ku, Sapporo 060-8628 Japan.

3. Department of Physical Science, Hiroshima University, 1-3-1 Kagamiyama, Higashihiroshima, Hiroshima 739-8526 Japan.

mizugu@tmu.ac.jp

**Abstract**. $Eu_{0.5}La_{0.5}FBiS_{2-x}Se_x$ is a new $BiS_2$-based superconductor system. In $Eu_{0.5}La_{0.5}FBiS_{2-x}Se_x$, electron carriers are doped to the $BiS_2$ layer by the substitution of Eu by La. Bulk superconductivity in this system is induced by increasing the in-plane chemical pressure, which is controlled by the Se concentration ($x$). In this study, we have analysed the crystal structure of $Eu_{0.5}La_{0.5}FBiS_{2-x}Se_x$ using synchrotron powder diffraction and the Rietveld refinement. The precise determination of the structural parameters and thermal factors suggest that the emergence of bulk superconductivity in $Eu_{0.5}La_{0.5}FBiS_{2-x}Se_x$ is achieved by the enhanced in-plane chemical pressure and the decrease in in-plane disorder.

## 1. Introduction

The $BiS_2$-based superconductors have been drawing much attention as a new layered superconductor family [1-3]. The typical crystal structure is tetragonal (*P4/nmm*) and is composed of alternate stacks of the electrically conducting $BiS_2$ layers and the insulating (blocking) layers, which resembles those of the Cu-based and the FeAs-based high transition temperature (high-$T_c$) superconductors [4,5]. The parent phase of the $BiS_2$-based compounds is a semiconductor with a band gap [1,6,7]. Upon the generation of the electron carriers in the $BiS_2$ layers, metallic conductivity is induced, and superconductivity is observed in the electron-doped compounds. With this scenario that the superconductivity can be induced by the electron carrier doping, many $BiS_2$-based superconductors have been discovered [3]. Recent studies, however, have suggested that the emergence of superconductivity in $BiS_2$-based compounds is related with not only the carrier doping but also the crystal structure optimization. For example, although the superconducting transitions are observed in $LaO_{1-x}F_xBiS_2$ when electron carriers were doped by substitution of F for O, bulk superconductivity never appears while filamentary superconductivity with a small shielding volume fraction is observed. In addition, $T_c$ in the filamentary superconducting phase is less than 3 K for optimally doped $x = 0.5$ [2]. The absence of bulk superconductivity in $LaO_{0.5}F_{0.5}BiS_2$ would be caused by the long bond length of in-plane Bi-S, in which orbital overlaps between Bi and S are small. According to our recent report, the enhancement of in-plane chemical pressure, which is related to the enhancement of orbital overlaps, are required to induce bulk superconductivity in the $LaO_{0.5}F_{0.5}BiS_2$ system [8].

Similar situation is observed in $Eu_{0.5}La_{0.5}FBiS_2$, which has a lattice structure similar to the $LaO_{1-x}F_xBiS_2$ system [9]. Although the electron carriers are doped by the substitution of Eu by La, bulk superconductivity is not observed. Instead, superconductivity is induced by the external pressure effect. $T_c$ is also enhanced to ~10 K by the application of external pressure [9]. The bulk superconductivity states of $BiS_2$-based compounds can be induced by not only the external pressure effect but also the chemical pressure effect. We demonstrated the Se substitution for the $Eu_{0.5}La_{0.5}FBiS_2$ system. Polycrystalline samples of $Eu_{0.5}La_{0.5}FBiS_{2-x}Se_x$ were synthesized up to $x = 1$, and the emergence of bulk superconductivity with a $T_c$ of ~ 4 K for $x \geq 0.6$ was confirmed by the electrical resistivity and magnetic susceptibility measurements [10]. From powder X-ray diffraction with the conventional laboratory X-ray source (a CuK$\alpha$ radiation) and the Rietveld refinements, we have suggested that the Se substitution resulted in the enhancement of the in-plane chemical pressure, and hence, the bulk superconductivity is induced. This result is consistent with previous studies on $LaO_{0.5}F_{0.5}BiS_{2-x}Se_x$ and $CeO_{0.5}F_{0.5}BiS_{2-x}Se_x$, in which the enhancement of in-plane chemical pressure induced bulk superconductivity [8,11,12]. To confirm that the in-plane chemical pressure scenario is universal to the $BiS_2$-based superconductors, precise analyses of the sample purity and crystal structure parameters of the newly discovered superconductor $Eu_{0.5}La_{0.5}FBiS_{2-x}Se_x$ are needed. In this study, we have performed synchrotron powder X-ray diffraction for $Eu_{0.5}La_{0.5}FBiS_{2-x}Se_x$. The evolution of crystal structure is analyzed by the Rietveld method.

## 2. Experimental details

The polycrystalline samples of $Eu_{0.5}La_{0.5}FBiS_{2-x}Se_x$ used in this study were prepared using a conventional solid-state-reaction method as described in Ref. 10. Synchrotron X-ray diffraction experiment was performed at the beam line BL02B2 of Spring-8 under a proposal No. 2016B1078. The synchrotron X-ray with an energy of 25 keV was used. Using the one-dimensional semiconductor detector (MYTHEN), diffraction data with a step of 0.006º were collected. All the experiments were performed at room temperature. The Rietveld refinements were performed using the RIETAN software [13]. For the in-plane Bi and chalcogen (Ch) sites, the Rietveld analyses were performed with the anisotropic thermal factors. The crystal structure image was drawn using the VESTA software [14].

## 3. Results and discussion

The X-ray diffraction patterns for $x = 0$–1 were refined using a tetragonal $P4/nmm$ space group. For $x = 0, 0.2, 0.4$, and $0.6$, fluoride impurities ($BiF_3$ and $LaF_3$) with populations of 4%, 4%, 5%, and 7% were found, respectively. For $x = 0.8$ and 1, small impurity peaks of the fluorides and unidentified broad peaks at $2\theta = 6.2º$ and $9.3º$ were observed. Although the broad impurity peaks would be selenides because of the appearance at higher nominal concentration of Se, we could not refine the impurity phase. Figure 1 displays the typical synchrotron X-ray diffraction pattern and Rietveld refinement fitting result for $x = 0.6$. Assuming the major phase ($x = 0.6$) and two fluoride impurities, the diffraction pattern is well fitted, and the resulting reliability factor ($R_{wp}$) is 8%.

With the obtained structural parameters, we discuss the evolution of crystal structure of Se-substituted $Eu_{0.5}La_{0.5}FBiS_{2-x}Se_x$. Figures 2(a) and 2(b) show the Se concentration dependences of lattice constant of $a$ and $c$. The lattice constant of $a$ monotonically increases with increasing $x$. In contrast, the lattice constant of $c$ does not change for $x = 0$–0.6 and slightly increases at $x \geq 0.8$. These results are consistent with the previous study performed with a laboratory X-ray diffractometer [10] and suggests that the Se is selectively occupies the in-plane Ch1 site (see the inset of Fig. 1 for the definition of the Ch1 and Ch2 site). As experimental facts, we have not succeeded the refinement of Se occupancy at the Ch1 and the Ch2 sites because of the resulting negative small value of Se occupancy at the Ch2 site, which may be due to the fact that almost 100% Se goes to the in-plane Ch1 site. Therefore, we did the Rietveld refinements with the fixed Se occupancy as the nominal values $x$.

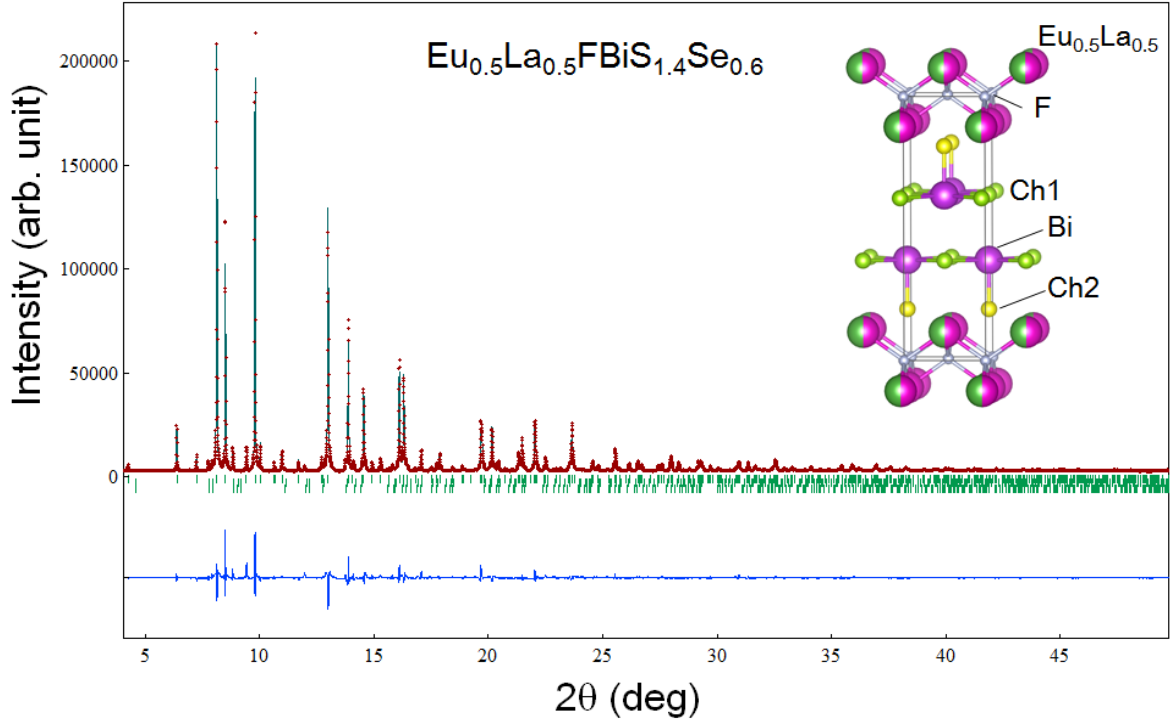

Fig. 1. X-ray diffraction pattern and the Rietveld fitting with a three-phase ($Eu_{0.5}La_{0.5}FBiS_{1.4}Se_{0.6}$, $BiF_3$, and $LaF_3$) analysis method for $Eu_{0.5}La_{0.5}FBiS_{1.4}Se_{0.6}$ ($x = 0.6$).

Figure 2(c) shows the Se concentration dependence of the Ch1-Bi-Ch1 bond angle. With increasing $x$, the bond angle decreases. Therefore, the flatness of the Bi-Ch1 plane becomes worse by Se substitution. This tendency is consistent with the structural evolution in the sister system $LaO_{0.5}F_{0.5}Bi(S,Se)_2$ [11]. Figures 2(d)–2(f) show the in-plane Bi-Ch1 bond distance, Bi-Ch2 bond distance, and inter-plane Bi-Ch1 bond distance. The evolution of the in-plane Bi-Ch1 distance seems to be correlating with the lattice constant of $a$. In contrast, the changes in the Bi-Ch2 distance and the inter-plane Bi-Ch1 distance seem to be correlating with the lattice constant of $c$. These direct correlation between the local structure parameters and the lattice constant also suggest the selective occupancy of Se at the in-plane Ch1 site.

As demonstrated in Refs. 8 and 10, the in-plane chemical pressure parameters ($CP$) were calculated using the obtained in-plane Bi-Ch1 distance and the ionic radii using the equation, $CP = (R_{Bi}$ and $R_{Ch}$/(in-plane Bi-Ch1 distance). $R_{Bi}$ is the ionic radius of $Bi^{2.5+}$, estimated in a previous structural analysis of a single crystal of $La(O,F)BiS_2$ [15]. $R_{Ch}$ (= 104.19 pm) is the average value of the ionic radius at the Ch1 site, which is calculated using nominal $x$ and ionic radii of S (184 pm) and Se (198 pm). With increasing Se concentration, the in-plane chemical pressure increases. Although the in-plane Bi-Ch1 distance increases by Se substitution (Fig. 2(d)), the in-plane packing density increases by Se substitution, which results in the enhancement of orbital overlap between Bi and Se. As proposed in Refs. 8 and 10, we have confirmed that the enhancement of the in-plane chemical pressure is essential for the emergence of bulk superconductivity in this system as well.

Finally, we discuss the evolution of the in-plane disorder by Se substitution. One of the advantages of synchrotron X-ray diffraction is the precise determination of the thermal factor, which can represent information about structural disorder. Since the superconductivity is induced in the Bi-Ch1 plane, we performed the Rietveld refinements with the anisotropic thermal factors for Bi and Ch1. Figure 2(h) shows the Se concentration dependences of the anisotropic thermal factor $U_{11}$ for in-plane Bi and Ch1 sites. $U_{11}$ for Bi does not show a remarkable changes with increasing $x$. In contrast, $U_{11}$ for Ch1 shows a notable dependence on $x$. With increasing Se concentration, $U_{11}$ for Ch1 decreases, which indicates that the in-plane disorder is suppressed by Se substitution. Generally in the BiS$_2$-based compounds, the existence of huge in-plane disorders has been detected using several probes (neutron diffraction, X-ray diffraction, and X-ray absorption spectroscopy) [16-20]. We consider that the effect of the in-plane chemical pressure on the emergence of superconductivity in Se-substituted Eu$_{0.5}$La$_{0.5}$FBiS$_{2-x}$Se$_x$ would be the suppression of the in-plane disorder. This scenario is consistent with the results of extended X-ray absorption fine structure studies on BiS$_2$ compounds, in which the in-plane Bi-S1 bond intensity is enhanced by the in-plane chemical pressure effect generated by the small RE (rare earth ion) substitution in REO$_{0.5}$F$_{0.5}$BiS$_2$ [8,20]. Therefore, with the present structural studies on the Eu$_{0.5}$La$_{0.5}$FBiS$_{2-x}$Se$_x$ system, we propose the relationship between the local in-plane disorder, the in-plane chemical pressure effect, and the emergence of superconductivity in the BiCh$_2$-based superconductor family. To further confirm this scenario, structural studies with local scale and/or single crystal studies are needed.

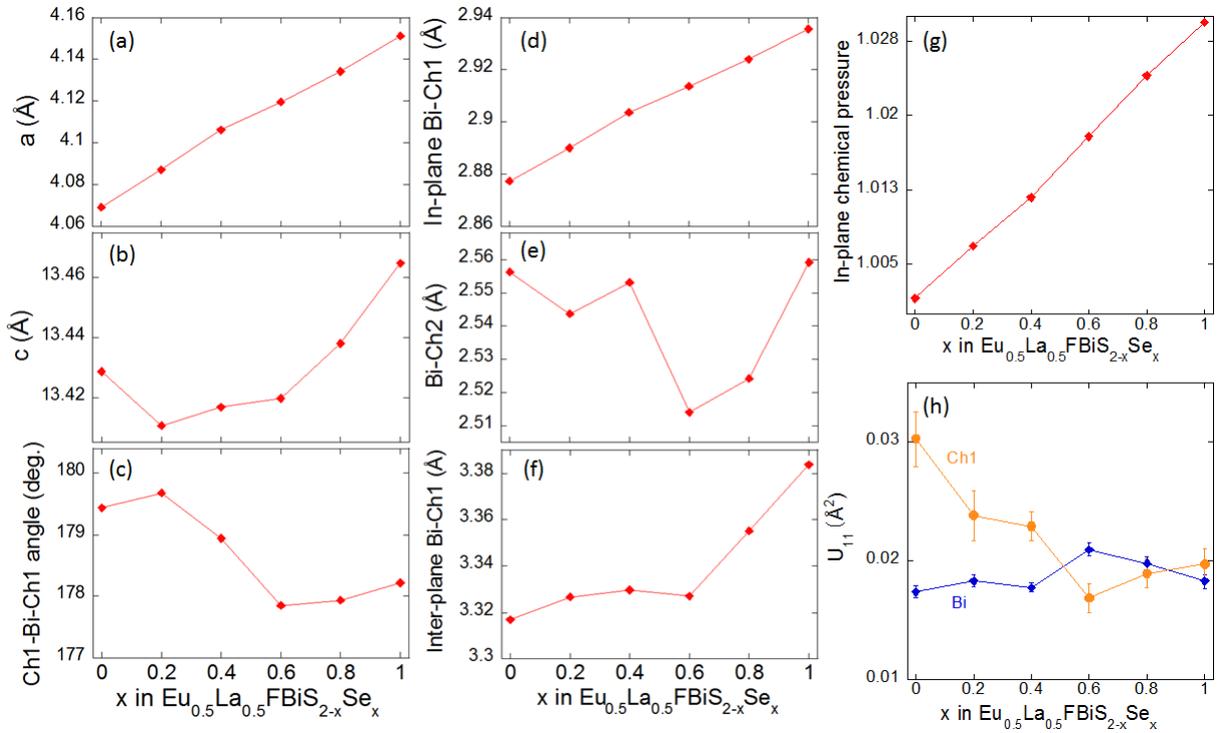

Fig. 2. Se concentration dependences of structural parameters for Eu$_{0.5}$La$_{0.5}$FBiS$_{2-x}$Se$_x$: (a) lattice constant of $a$, (b) lattice constant of $c$, (c) Ch1-Bi-Ch1 bond angle, (d) in-plane Bi-Ch1 bond distance, (e) Bi-Ch2 bond distance, (f) inter-plane Bi-Ch1 bond distance, (g) in-plane chemical pressure, and (h) anisotropic thermal factor of $U_{11}$ for in-plane Bi and Ch1 sites.

## 4. Summary

In this study, we have performed synchrotron X-ray powder diffraction for the new BiS2-based superconductor system Eu$_{0.5}$La$_{0.5}$FBiS$_{2-x}$Se$_x$, in which the bulk superconductivity was induced by Se substitution. The obtained structural parameters suggest four structural tendencies: (1) the selective occupancy of Se at the in-plane Ch1 site, (2) the direct correlation between Bi-Ch bond distances and the lattice constants, (3) the enhancement of the in-plane chemical pressure, and (4) the suppression of the in-plane disorder by Se substitution. These results will be useful for understanding the relationship between superconductivity and crystal structure in the BiCh$_2$-based superconductor family.

**Acknowledgements**
The authors would like to thank the staffs of BL02B2 of SPring-8 for the experimental support. This work was partly supported by Grants-in-Aid for Scientific Research (Nos. 16H04493 and 15H05886) and SPring-8 (proposal No. 2016B1078).